\documentclass{elsart}

\usepackage{epsfig}

\usepackage{amssymb}

\begin{document}

\begin{frontmatter}

\title{An analysis on extrema and constrained bounds for the soft 
Pomeron intercept}
\author{E. G. S. Luna, M. J. Menon and J. Montanha}
\address{Instituto de F\'{\i}sica Gleb Wataghin, \\
Universidade Estadual de Campinas,
13083-970 Campinas, SP, Brazil}

\date{\today}

\begin{abstract}
We investigate some aspects and consequences of the extrema 
bounds for the soft Pomeron intercept, recently determined 
by means of global fits to $pp$ and $\bar{p}p$ total cross 
section data at both accelerator and cosmic-ray energy regions
(scattering data). 
We also examine the effects of the secondary Reggeons by 
introducing fitted trajectories from Chew-Frautschi plots
(spectroscopy data)
 and 
determining new constrained bounds for the Pomeron intercept. 
In both cases we extend 
the analysis to $baryon-p$, $meson-p$, $baryon-n$, $meson-n$,
$gamma-p$ and 
$gamma-gamma$ scattering, presenting tests on factorization 
and quark counting rules. We show that in all the cases 
investigated, the bounds lead to good descriptions of the 
bulk of experimental data on the total cross sections, 
but with different extrapolations to higher energies. Our main 
conclusion is that the experimental information presently 
available on the above quantities is not sensitive to an 
uncertainty of 2\% in the value of the soft Pomeron intercept.
At 14 TeV (CERN LHC) the extrema and constrained bounds allow
to infer $\sigma_{tot}$ = 114 $\pm$ 25 mb and 105 $\pm$ 10 mb,
respectively.
\end{abstract}

\begin{keyword}
elastic hadron scattering \sep total cross sections \sep Regge formalism 

\PACS 13.85.Dz \sep 13.85.Lg \sep 11.55.Jy
\end{keyword}
\end{frontmatter}

\section{Introduction}

The total cross section is one of the most important physical quantities
that characterizes the hadron-hadron scattering. From Unitarity (optical
theorem) this quantity is expressed in terms of the {\em forward}
elastic scattering amplitude and at high energies the relation reads

\begin{equation}
\sigma_{tot}(s) = \frac{\textrm{Im}\ F(s,t=0)}{s},
\label{sigtot}
\end{equation}
where $s$ is the center-of-mass energy squared and
$t$ is the four momentum transfer squared.
From the experimental point of view, the general smooth increase of all
hadronic total cross sections, above $\sqrt s \approx $ 20 GeV, is a
well established result \cite{pdg}. However, since this rise is inherently 
a nonperturbative phenomena (soft diffractive process), the theoretical
treatment is still essentially phenomenological. 

Models rely on
general principles of Quantum Field Theory and, among them, the
Regge Pole formalism plays a fundamental role \cite{ddln}.
In this context, the elastic scattering amplitude in the $s$-channel is
expressed as a descending {\em asymptotic} series of powers
of $s$, each term representing a specific exchange in the
$t$-channel \cite{ddln,bp}

\begin{equation}
F(s, t) = \sum_{k} \gamma_{k}(t) \zeta_{k}(t) s^{\alpha_k (t)},
\end{equation}
where $\gamma_{k}(t)$ is the residue function, $\zeta_{k}(t)$ is
the (complex) signature factor and $\alpha_k(t)$ the trajectory function.

The decreasing of the total cross sections below 
 $\sqrt s \approx $ 20 GeV is assumed to be a consequence
of the exchange of meson resonances families, the Reggeons ($\tt I\!R$),
with the adequate quantum numbers in the $t$-channel process,
and represented by trajectories interpolating the data on
plots of spin $J$ versus the square of their masses (Chew-Frautschi plot).
That scheme provides trajectories that are approximately linear in $t$,
$\alpha_{\tt I\!R}(t) \approx 
\alpha_{\tt I\!R}(0) +
\alpha_{\tt I\!R}^{'} t$, and for the known mesonic resonances we have
intercepts $\alpha_{\tt I\!R}(0) \approx 0.5$ and
slopes $\alpha_{\tt I\!R}^{'} \approx 1.0$ GeV$^{-2}$. From Eqs. (1) and
(2), and denoting $\gamma_{k}(t=0)\ \mathrm{Im}\ \zeta_{k}(t=0) \equiv g_{\tt I\!R_k}$
(the Reggeon strength), the total cross section reads

\begin{equation}
\sigma_{tot}(s) = \sum_{k} g_{\tt I\!R_k} s^{\alpha_{\tt I\!R_k}(0) -1},
\end{equation}
and therefore, decreases roughly as $1/ \sqrt s$.
On the other hand, the rise of the total cross sections above
$\sqrt s \approx $ 20 GeV is phenomenologically implemented by
the introduction of an \textit{ad hoc} trajectory, associated with
a colourless state having the vacuum quantum numbers, the Pomeron ($\tt I\!P$),
with intercept
slightly greater than one: $\alpha_{\tt I\!P}(t) =
\alpha_{\tt I\!P}(0) +
\alpha_{\tt I\!P}^{'} t$, $\alpha_{\tt I\!P}(0) >  1$
(Supercritical Pomeron). 
Since there are no known particles related with the Pomeron
trajectory (except for the glueball candidate $2^{++}$ \cite{land01}), 
it is necessary to perform fits to the available data in
order to establish the Pomeron's parameters, namely, its intercept 
$\alpha_{\tt I\!P}(t=0)$ and strenght $g_{\tt I\!P}$. 

Donnachie and
Landshoff have shown that the total cross sections on $pp$, $\bar{p}p$,
$meson-p$, $\gamma-p$, $pn$, and $\bar{p}n$
 scattering can be well described with the Pomeron and
a degenerate Reggeon contribution (associated with the 
$a_2, f_2, \rho,$ and $\omega$
families of particles) \cite{dl92}. A remarkable fact is the result that all 
the data show agreement with the same value for the soft Pomeron intercept, namely 
$\alpha_{\tt I\!P}(t=0) = 1.0808$. This means that the rise of the total
cross sections is a direct consequence of the ``object" exchanged, the Pomeron, and
does not depend on the intrinsic structure of the hadrons involved in
the scattering.

More recently, new data analyzes with extended models that allow
the splitting of the Reggeon trajectories (non-degenerate $C = +1$ and
$C = -1$ meson trajectories) or using different estimations for
$\sigma_{tot}$ from cosmic-ray experiments,
have indicated larger values for the
Pomeron intercept, namely 1.09 - 1.12 \cite{cmg,ckk,compete,alm03}. In
general these analyzes
are characterized by global fits of the Pomeron/Reggeon parameters to the
forward scattering data, that is, without using as input the fitted
trajectories from the Chew-Frautschi plots.

However, independently of the parametrization (model) used, to select
a correct or ``secure"  value for the intercept is a difficult task, 
mainly due
to the well known disagreement between measurements of the $\bar{p}p$ total cross
sections at the highest energy reached in accelerators, $\sqrt s =$ 1.8 TeV
(CDF and E710/E811 results, \cite{cdf,e710/811}) . Experimental information on $pp$ total 
cross sections from cosmic-ray experiments exist in the
energy region
$\sqrt s :$ 6 - 40 TeV, but they are also characterized by discrepancies and 
large uncertainties \cite{alm03}.

Therefore, despite the fundamental role of the soft Pomeron intercept in the
investigation of the rise of the hadron-hadron total cross sections,
its ``exact" or accepted value still remains an open problem and that is a
consequence of the uncertainties in the experimental information at the highest
energies.

Based on that fact, in a previous work we have developed a quantitative
investigation of the effect of the discrepant data/information on the
value of the Pomeron intercept \cite{lm03}. By combining the highest
or lowest results for the $pp$ and $\bar{p}p$ total cross sections from
both accelerator and cosmic-ray experiments, and testing all the important
variants of fits
to the experimental data above 10 GeV through an extended Regge
parametrization (non-degenerate trajectories), we have determined extrema
upper and lower bounds for the intercept,
1.109 and 1.081, respectively. That permits to infer the 
fastest and slowest
increase scenarios for the rise of the hadronic total cross sections,
allowed by the experimental information presently available.

In this work, we investigate the effects of these bounds in a global study
of the hadron-hadron elastic scattering. For each fixed 
\textit{extrema bound}, previously
determined from $pp$ and $\bar{p}p$ scattering, we extend the analysis
to $p^{\pm}n$, $\pi^{\pm}p$, $K^{\pm}p$, $K^{\pm}n$, $\Sigma^{-}p$, 
$\gamma p$ and $\gamma \gamma$ 
elastic scattering (here $p^{\pm}$ indicates $p$ and $\bar{p}$), 
by means of simultaneous fits to the 
corresponding total cross section data
(some partial results have already been presented in
\cite{lmm04bjp}). In addition, using as input the 
fitted secondary trajectories from the Chew-Frautschi plots,
we determine new
\textit{constrained bounds} for the intercept from $pp$ and $\bar{p}p$
scattering. These results are also extended to the above
reactions. In all the cases investigated we present tests on factorization 
rule associated with
$pp$, $\gamma p$, and $\gamma \gamma$ scattering and quark counting rule
in $p^{\pm}p$, $p^{\pm}n$, $\pi^{\pm}p$, $K^{\pm}p$, $K^{\pm}n$, and 
$\Sigma^{-}p$ scattering.
We show that both upper and lower extrema bounds for the intercept lead to good 
descriptions of the experimental data and information presently available. 
However, 
extrapolations to higher
energies indicate different scenarios for the rise of the total cross sections.

The paper is organized as follows. In Section 2 we treat the bounds for the
intercept from analyzes of $pp$ and $\bar{p}p$ data, by means of both
global fits to the total cross section data (extrema bounds) and also using
as input fitted secondary Reggeon trajectories (constrained bounds).
In Section 3 we present the extensions to $p^{\pm}n$,
$\pi^{\pm}p$, $K^{\pm}p$, $K^{\pm}n$, $\Sigma^{-}p$, $\gamma p$, and $\gamma \gamma$ 
scattering and tests concerning factorization and quark-counting rules.
The conclusions and some final remarks are the contents of Section 4.

\section{\label{Sect:2} Bounds for the Pomeron intercept from $p p$ and 
$\bar{p} p$ scattering}

In this Section we first introduce the extended parametrization
(non-degenerate meson trajectories) to be used in all the fits.
Afterwards we 
shortly review the results and notation used
in the previous determination of the {\em extrema bounds} for the Pomeron
intercept (scattering data), and then discuss the determination of 
{\em constrained bounds}
obtained with fitted secondary Reggeon trajectories
(spectroscopy data). We end the Section with a comparison between
the results for the $pp$ and $\bar{p}p$ total cross sections
obtained with both extrema and constrained bounds.

\subsection{Extended Regge parametrization}

In the extended  Regge Pole model, the forward scattering 
amplitude is decomposed into
three contributions \cite{cmg,ckk}: 
$F(s) = F_{\tt I\!P}(s) + F_{a_2/f_2}(s) +  \tau F_{\rho/\omega}(s)$,
where
$F_{\tt I\!P}$ represents the single Pomeron,
$F_{a_2/f_2}$ the Reggeon with $C=+1$,
$F_{\rho/\omega}$ the Reggeon with $C=-1$,
and $\tau = + 1$ ($- 1$) for antiparticle-particle (particle-particle)
scattering.
The intercepts of the Pomeron, the $C=+1$, and the $C=-1$ 
trajectories are expressed by
$\alpha_{\tt I\!P}(0)  =  1+\epsilon$,
$\alpha_{+}(0)  =  1 -\eta_{+}$, and
$\alpha_{-}(0)  =  1 -\eta_{-}$, respectively.
With this amplitude, and through the optical theorem,
Eq. (\ref{sigtot}), the total cross section for 
particle-particle and
antiparticle-particle
interactions reads
\begin{equation}
\sigma_{tot}(s) = X s^{\epsilon} + Y_{+}\, s^{-\eta_{+}} + \tau Y_{-}\, 
s^{-\eta_{-}},
\label{eq:total}
\end{equation}
where the coefficients $X$, $Y_{+}$, $Y_{-}$, and the exponents $\epsilon$, 
$\eta_{+}$, and $\eta_{-}$ are
parameters to be determined by the fit to the data.

Two shortcomings must be recalled about this model. One refers to the intrinsic
asymptotic character of Eq. (4), since it is intended for the region
$s/s_0 \rightarrow \infty$, with $s_0 \approx$ 1 GeV. Therefore one must
be careful to infer physical meanings at finite energies. The other concerns
the violation of the Froissart-Martin bound by power laws with exponents greater
than one, as it is the case for the Supercritical Pomeron. However, it is 
understood that $\epsilon$ is
an effective parameter, which may eventually depend on the energy. In spite of
these shortcomings, the Regge formalism constitutes presently one of the most
useful bridges to a well founded theoretical approach (non-perturbative
QCD) for soft diffractive processes \cite{ddln}.

\subsection{Extrema bounds}

The extrema bounds for the Pomeron intercept have been determined in
Ref. \cite{lm03} from analyzes of the discrepancies in 
$pp$ and $\bar{p}p$ total cross section data at both accelerator
and cosmic-ray energy regions. For $\bar{p}p$ scattering the discrepancies
appear in the values of $\sigma_{tot}^{\bar{p}p}$ at $\sqrt s$ = 1.8 TeV.
The highest value concerns the measurement by the CDF Collaboration (CDF)
\cite{cdf}
and the lowest values, the measurements by the E710 and E811 Collaborations
(E710/E811) \cite{e710/811}. In the case of $pp$ total cross sections, 
extracted from
$p$-air cross sections (cosmic rays) at $\sqrt s:$ 6 - 40 TeV, the
highest estimations concern the results by Nikolaev and also by
Gaisser, Sukhatme and Yodh (NGSY) \cite{ngsy} and the lowest estimations,
the results by Block, Halzen and Stanev (BHS)
\cite{bhs} (detailed discussion on the 
experimental uncertainties, together with 
numerical tables may be found in \cite{alm03}).

The strategy and method used in \cite{lm03} was the following. 
First, the {\it accelerator} data
on $pp$ and $\bar{p}p$ scattering were split in 
two ensembles, where each ensemble displays one of the two possible
scenarios for the total cross section behavior with energy:

Ensemble I -  $\sigma_{tot}^{pp}$ 
and $\sigma_{tot}^{\bar{p}p}$
data ($10 \leq \sqrt{s} \leq 900 \, GeV$) + CDF datum ($\sqrt{s} =
1800 \, GeV$);

Ensemble II - $\sigma_{tot}^{pp}$ 
and $\sigma_{tot}^{\bar{p}p}$
data ($10 \leq \sqrt{s} \leq 900 \, GeV$) + E710/E811 data 
($\sqrt{s} = 1800 \, GeV$).

The choice for the minimal energy at $10$ GeV was based on an
analysis showing that the parameters of Regge fits are stable for
a cutoff at $\sqrt s \approx 9$ GeV \cite{compete}. We shall return
to that point at the end of this Subsection.

In a second step, since with the extended Regge parametrization (4)
one has
$\sigma_{tot}^{\bar{p}p}(s) - \sigma_{tot}^{pp}(s) \rightarrow$ 0
as $s \rightarrow \infty$, 
the highest and lowest estimations for $\sigma_{tot}^{pp}$
from {\it cosmic-ray} experiments have been adequately added
to the above ensembles, defining two other cases: 

Ensemble I + NGSY (fastest increase scenario); 

Ensemble II + BHS (slowest increase scenario).

Besides individual fits to the total cross sections, the
available data on the ratio of the real to the imaginary part 
of the amplitude, $\rho(s) = \textrm{Re} \ F(s,t=0) / \textrm{Im} 
\ F(s,t=0)$,   
were also used in global fits involving $\rho$ and $\sigma_{tot}$.
This has been done by means of dispertion relations and either using
the subtraction constant $K$ as a free fit parameter or assuming
$K =$ 0. 
Among the 16 variants of fits performed through the CERN-Minuit routine,
the \textit{extrema values} for the Pomeron intercept were obtained in the case
of individual fits to $\sigma_{tot}$, the highest value with 
Ensemble I + NGSY and the lowest one with Ensemble II,
$\alpha_{\tt I\!P}^{upper}(0) = 1.104 \pm 0.005$ and
 $\alpha_{\tt I\!P}^{lower}(0) = 1.085 \pm 0.004$, respectively.
The \textit{extrema upper bound} was inferred by adding the corresponding error
to the central upper value and the \textit{extrema lower bound} by subtracting
the corresponding error from the central lower value: $1.109$ and $1.081$, respectively.
The results of the fitting are displayed in Table \ref{tab:1}, together
with the extrema bounds for the parameter $\epsilon$.

\begin{table}[h]
\begin{center}
\caption{\textit{Extrema} values and bounds for the Pomeron 
intercept ($\alpha_{\tt I\!P}(0) = 1 + \epsilon$):
parameters obtained through global fits to $pp$ and $\bar{p}p$
total cross section data \cite{lm03}.}
\label{tab:1}
\begin{tabular}{ccc}
\hline
Ensemble: & I + NGSY & II  \\ 
 & (upper) & (lower) \\ \hline
$\epsilon_{\mathrm{extrema}}^{\mathrm{values}}$ & 0.104$\pm$0.005 
& 0.085$\pm$0.004  \\
$X$ (mb) & 16.4$\pm$1.2 & 20.47$\pm$0.88 \\
$\eta_{+}$ & 0.28$\pm$0.03 & 0.38$\pm$0.04  \\
$Y_{+}$ (mb) & 51.2$\pm$4.2 & 62.2$\pm$7.5 \\
$\eta_{-}$ & 0.42$\pm$0.04 & 0.42$\pm$0.04 \\
$Y_{-}$ (mb) & 17.4$\pm$3.8 & 17.2$\pm$3.9  \\
$\textrm{No.}\ F$ & 94 & 89  \\
$\chi^2 /F$ & 1.01 & 0.94 \\ 
$\epsilon_{\mathrm{extrema}}^{\mathrm{bounds}}$ & 0.109 & 0.081 \\
\hline
\end{tabular}
\end{center}
\end{table}

In that analysis, the secondary Reggeon intercepts
were left as free parameters in order to allow the necessary freedom for 
the Pomeron intercept 
to find its extrema. The final result for the Reggeon intercepts, from Table 
\ref{tab:1}, are:
\begin{eqnarray}
\alpha_{a_2/f_2}(0) & = & 1 - \eta_{+} = 0.72 \pm 0.03,  \nonumber \\
\alpha_{\rho/\omega}(0) & = & 1 - \eta_{-} = 0.58 \pm 0.04,
\end{eqnarray}
for the upper bound, and
\begin{eqnarray}
\alpha_{a_2/f_2}(0) & = & 1 - \eta_{+} = 0.62 \pm 0.04,     \nonumber \\
\alpha_{\rho/\omega}(0) & = & 1 - \eta_{-} = 0.58 \pm 0.04, 
\end{eqnarray}
 for the lower one. We shall return to these results in the next
Subsection.

In the present paper, in order to check the influence of the minimal energy 
(cutoff) in the determination of the
extrema values of the intercept, we performed several fits to the total
cross section with ensembles I + NGSY and II and varying
$\sqrt s_{\mathrm{min}}$ around $10$ GeV. The results for the upper and
lower values of the parameter $\epsilon$ are displayed in Fig. 1, together 
with the
$\chi^2/F$ obtained in each case. We see that although the lower values
indicate more stability than the upper values, the cutoff at $10$ GeV
provide good statistical results (in accordance with the results 
obtained in \cite{compete}).

\begin{figure}[ht]
\begin{center}
\includegraphics[width=14.0cm,height=10cm]{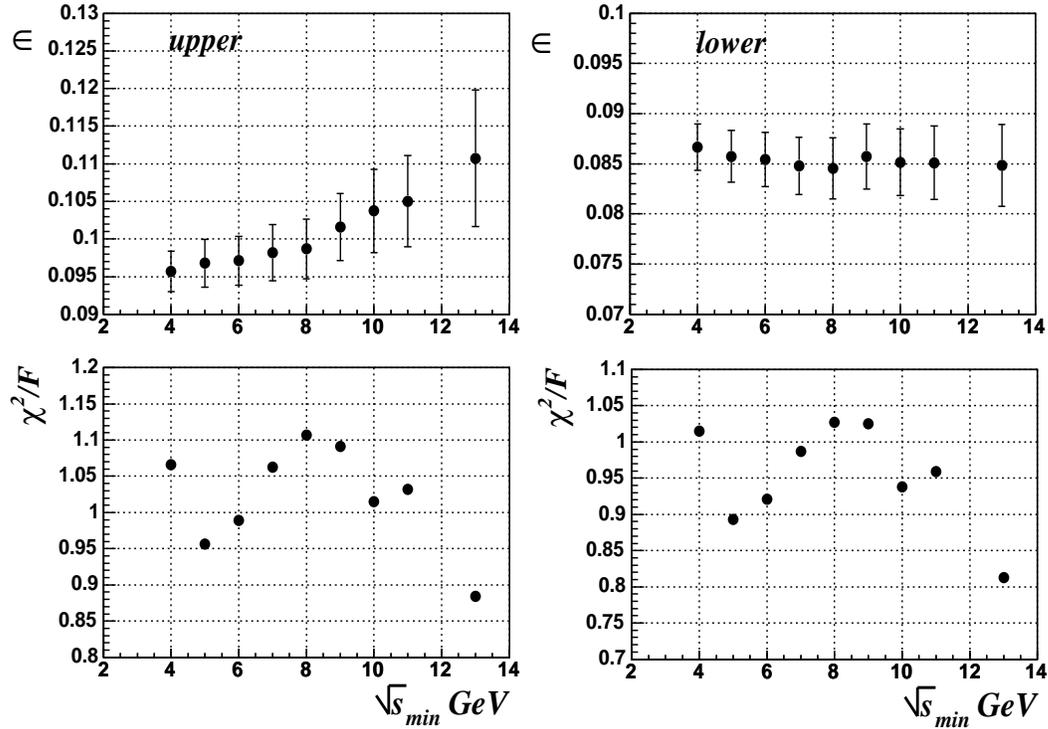}
\end{center}
\caption{Behavior of the upper and lower \textit{extrema values} 
of the parameter $\epsilon = \alpha_{\tt I\!P}(0) - 1$
(top) and the corresponding 
$\chi^{2}/F$ (botton) as function of the minimal energy
(cutoff), when the secondary 
Reggeon intercepts are left as 
free fit parameters to the scattering data (Table 1).}
\label{fig:1}
\end{figure}

\subsection{ Constrained bounds}

As mentioned in our introduction, the secondary Reggeon intercepts can also 
be determined directly from the Chew-Frautschi plots of spin versus squared mass
(spectroscopy data).
In what follows we show that using these intercepts as inputs we restrict
the upper and lower values and bounds of the Pomeron intercept.
To that end, we gathered the available data of
meson resonance families $a_2$, $f_2$, $\rho$ and $\omega$ in a Chew-Frautschi
plot, as shown in Figure 2. A fit of a linear expression to the
combined $a_2/f_2$ and $\rho/\omega$ data provided the trajectories
\begin{eqnarray}
\label{Reggeon1}
\alpha_{a_2/f_2} (t) & = & (0.548 \pm 0.016) + (0.847 \pm 0.009)\,t \ \ \ 
(\chi^2/F = 258), \\ 
\label{Reggeon2}
\alpha_{\rho/\omega} (t) & = & (0.442 \pm 0.003)+(0.912 \pm 0.005)\,t \ \ \ 
(\chi^2/F = 63.7),
\end{eqnarray}
which are also displayed in Fig. 2.

\begin{figure}[ht]
\label{fig:2}
\begin{center}
\includegraphics[width=10.0cm,height=8cm]{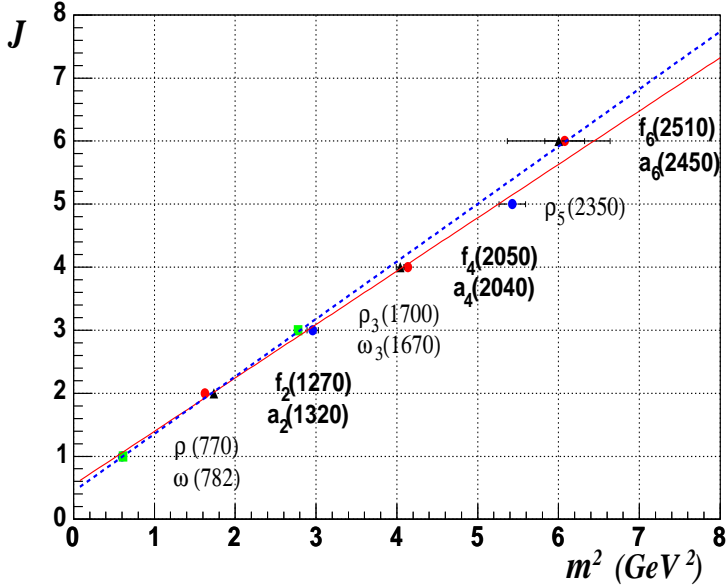}
\end{center}
\caption{Chew-Frautschi plot of the $a_2$, $f_2$, $\rho$, and $\omega$ resonance 
families
and the corresponding fitted trajectories for $a_2/f_2$ (solid line) 
and $\rho/\omega$
(dashed line), Eqs. (7) and (8), respectively.}
\end{figure}

For individual trajectories, the best fit provided
\begin{eqnarray}
\label{Reggeon3}
\alpha_{a_2}(t) & = & (0.491   \pm 0.034)+(0.869 \pm 0.019)\,t  \ \ \ \ 
(\chi^2/F = 0.3),  \\
\label{Reggeon4}
\alpha_{f_2}(t) & = & (0.676  \pm 0.017)+(0.814 \pm 0.010)\,t  \ \ \ \ 
(\chi^2/F = 3.8),   \\
\label{Reggeon5}
\alpha_{\rho}(t) & = & (0.501 \pm 0.007)+(0.840 \pm 0.012)\,t \ \ \ \ 
 (\chi^2/F = 0.2), \\
\label{Reggeon6}
\alpha_{\omega}(t) & = &(0.435)+(0.923)\,t \ \ \ \  
(\mathrm{not\ fitted}).
\end{eqnarray}
Overall, these results show some agreement with previous analysis by 
Desgrolard \textit{et al}.
\cite{Desgrolard}. 

From these fits, it is possible to establish a range for the $C=+1$ and $C=-1$
Reggeon intercept values, with Eqs. (\ref{Reggeon1}) and  (\ref{Reggeon2}) 
providing the central
values and Eqs. (\ref{Reggeon3}-\ref{Reggeon6}) providing the upper and lower 
bounds for each
set. From that, we have:
\begin{eqnarray}
\label{Reggeon7}
\alpha_{a_2/f_2}(0)  =  1 - \eta_{+} = 0.548^{+0.145}_{-0.091}, \\  \nonumber \\ 
\label{Reggeon8}
\alpha_{\rho/\omega}(0)  =  1 - \eta_{-} = 0.442^{+0.066}_{-0.007}.
\end{eqnarray}
These results can be compared with those obtained from fits to scattering data
in the determination of the upper and lower extrema bounds, Eqs. (5) and
(6), respectively. For the $a_2/f_2$ trajectory, the Reggeon intercept
in the case of the lower bound is compatible with the above spectroscopy
result and for the upper bound the value is barely above the range of Eq. (13).
For the $\rho/\omega$ trajectory the results with both upper and lower
bounds are above the range of Eq. (14). These discrepancies may be associated
with the particular variant that provided the extrema values for the
Pomeron intercept in Ref. \cite{lm03}, that is, fits to the total cross
section data only. As demonstrated in that paper, the inclusion of the $\rho(s)$
data constrains the asymptotic rise of the total cross section,
restraining the evaluation of maxima bounds.

In order to investigate the effect that extracting the secondary Reggeon 
intercepts from spectroscopy data can 
have on the value of the
Pomeron intercept, we performed new fits with both Reggeon intercepts fixed 
at their 
central
values given by Eqs. (\ref{Reggeon7}) and (\ref{Reggeon8}) and
letting free all the other parameters in Eq. (4). 
We used the same two sets of data that provided the extrema bounds of Table 
\ref{tab:1}, i.e.,
Ensemble I + NGSY
and Ensemble II, and $\sqrt s_{\mathrm{min}} = 10$ GeV (see below).
The results for these new fits are displayed in Table \ref{tab:2}. 
As in the case of the extrema
bounds, we can infer here the \textit{constrained bounds} by adding and 
subtracting
the uncertainties to the corresponding upper and lower
central values, respectively (also displayed in Table \ref{tab:2}).

\begin{table}[h]
\begin{center}
\caption{\textit{Constrained} values and bounds for the
Pomeron intercept ($\alpha_{\tt I\!P}(0) = 1 + \epsilon$):
parameters obtained from fits to $pp$ and $\bar{p}p$
total cross section with fixed secondary Reggeon intercepts
(central values in Eqs. (13) and (14)).}
\label{tab:2}
\begin{tabular}{ccc}
\hline
Ensemble: & I + NGSY & II  \\ 
 & (upper) & (lower) \\ \hline
$\epsilon_{\mathrm{constrained}}^{\mathrm{values}}$ & 0.087$\pm$0.002
 & 0.082$\pm$0.002  \\
$X$ (mb) & 20.49$\pm$0.30 & 21.30$\pm$0.29 \\
$\eta_{+}$ & 0.452 (fixed) & 0.452 (fixed)  \\
$Y_{+}$ (mb) & 86.3$\pm$2.2 & 81.2$\pm$2.1 \\
$\eta_{-}$ & 0.558 (fixed) & 0.558 (fixed) \\
$Y_{-}$ (mb) & 36.22$\pm$0.76 & 36.17$\pm$0.76  \\
$\textrm{No.}\ F$ & 96 & 91  \\
$\chi^2 /F$ & 1.26 & 1.03 \\
$\epsilon_{\mathrm{constrained}}^{\mathrm{bounds}}$ & 0.089 & 0.080 \\
\hline
\end{tabular}
\end{center}
\end{table}

As in the previous subsection, we have also checked the influence of the
minimal energy on the constrained values of the intercept. The results are
displayed in Fig. 3 showing, once more, that the stability in the intercepts
and good statistical results are obtained for the cutoff at
$\sqrt s_{\mathrm{min}} = 10$ GeV.

\begin{figure}[ht]
\begin{center}
\includegraphics[width=14.0cm,height=10cm]{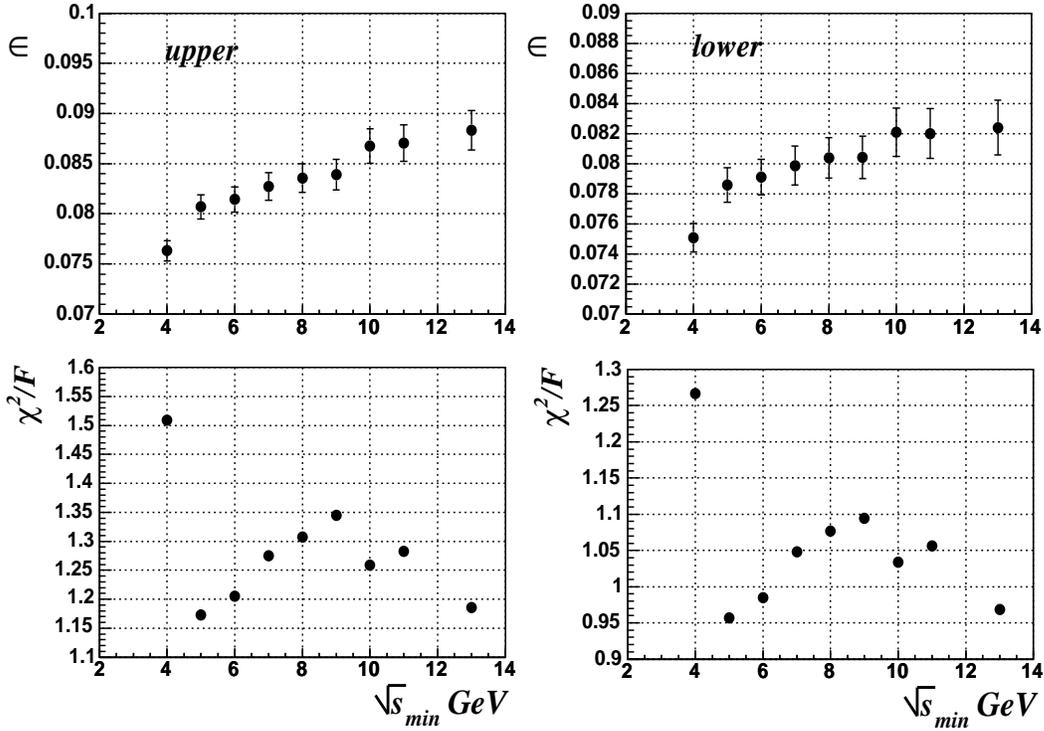}
\end{center}
\caption{Behavior of the upper and lower \textit{constrained values} 
of the parameter $\epsilon = \alpha_{\tt I\!P}(0) - 1$
(top) and the
corresponding $\chi^{2}/F$ (botton) as function of the minimal energy, 
when the secondary Reggeon intercepts are fitted to the spectroscopy
data (Table 2).}
\label{fig:3}
\end{figure}

\subsection{Bounds for the rise of the $pp$ and $\bar{p}p$ total 
cross sections}

The results for the $pp$ and $\bar{p}p$ total cross sections obtained by
means of the extended Regge parametrization (4) with
both the \textit{extrema} and \textit{constrained} bounds 
(Tables 1 and 2) are displayed in Fig. 4, 
together with the experimental data and estimations
used in the ensembles (see \cite{alm03} and 
\cite{lm03}
for references and discussions on these and others numerical values).

\begin{figure}[ht]
\begin{center}
\includegraphics[width=12.0cm,height=10cm]{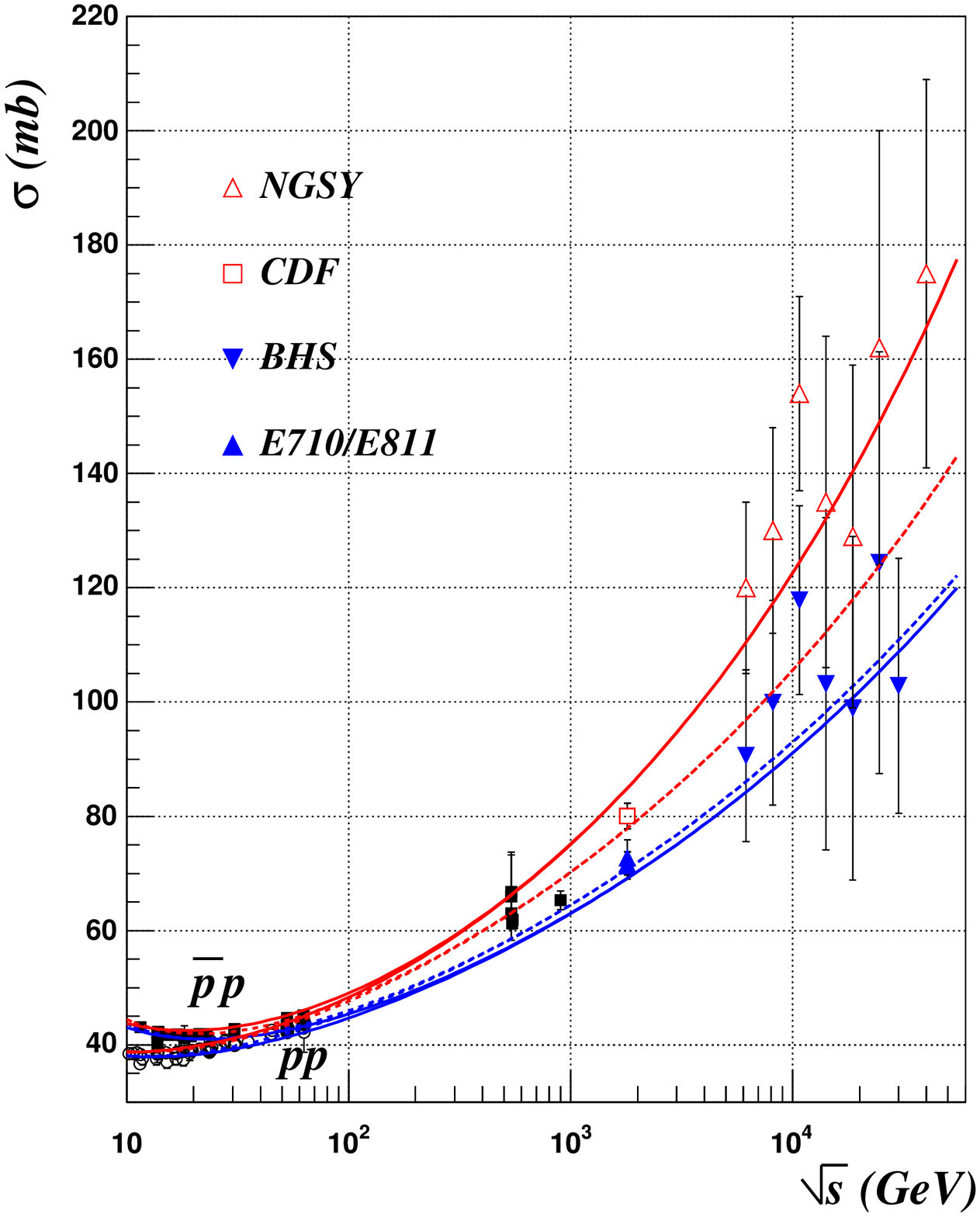}
\end{center}
\caption{Scenarios for the rise of the total cross sections from parametrization
(\ref{eq:total}), with the \textit{extrema} bounds (solid) and
the \textit{constrained} bounds (dashed), and parameters from
Tables \ref{tab:1} and \ref{tab:2}, respectively
(see \cite{alm03} and \cite{lm03} for discussions on the numerical points 
displayed at the cosmic-ray energy region).}
\label{fig:4}
\end{figure}

From Tables 1 and 2, we see that the best statistical results ($\chi^2/F$) were
obtained with the extrema bounds and from the curves in Fig. 4 we notice
the following aspects: (1) fixing the secondary Reggeon
intercepts to the spectroscopy data (non-degenerate $a_2/f_2$ and $\rho/\omega$
with linear parametrization) curbs the rising of the total cross sections
by limiting the freedom of the Pomeron intercept; (2) the region delimited
by the constrained bounds favors the E710/E811 and BHS
results, indicating a ``conservative" scenario; (3) the curve obtained
with the extreme upper bound is compatible with the NGSY result but is above 
the CDF result; (4) from the regions delimited by the upper and lower
extrema and constrained bounds we can infer averaged values for the total
cross sections at the BNL RHIC and CERN LHC energies,
200 GeV and 14 TeV, respectively:

\begin{eqnarray}
\sigma_{\mathrm{tot}}^{\mathrm{constrained}}(200\ \mathrm{GeV})
= 51.8 \pm 2.1\ \mathrm{mb},
\quad
\sigma_{\mathrm{tot}}^{\mathrm{extrema}}(200\ \mathrm{GeV})
= 51.9 \pm 3.8\ \mathrm{mb}, \nonumber
\end{eqnarray}

\begin{eqnarray}
\sigma_{\mathrm{tot}}^{\mathrm{constrained}}(14\ \mathrm{TeV})
= 105 \pm 10\ \mathrm{mb},
\quad
\sigma_{\mathrm{tot}}^{\mathrm{extrema}}(14\ \mathrm{TeV})
= 114 \pm 25\ \mathrm{mb}. \nonumber
\end{eqnarray}
We notice that the above range provided by the constrained bound at
14 TeV is in agreement with the QCD-inspired model prediction by 
Block, Halzen, and Stanev, namely $\sigma_{tot} = 108.0 \pm 3.4$ mb
\cite{bhs}.

\section{\label{Sect:3}Extensions to $p^{\pm} n$, $\pi^{\pm} p$, $K^{\pm} p$,
 $K^{\pm} n$, $\Sigma^{-}p$, $\gamma p$ and $\gamma \gamma$ scattering}

Besides $pp$ and $\bar{p}p$, several other hadronic reactions have 
been measured
through the last decades \cite{pdg}. Although none of them has the 
energy range of 
$pp$ and $\bar{p}p$, some reactions have been measured up to 
considerably high energy values.
For meson-proton, the total cross section data for 
$K^{\pm}p$ have been recorded up to $\sqrt s =$ 24.1 GeV, whereas for
$\pi^{\pm} p$ the top energy is $\sqrt s =$ 34.7 GeV.
For baryon-proton, the $\Sigma^{-} p$ has been measured up to
this same energy, whereas for $\gamma p$ and $\gamma \gamma$ there are 
measurements reaching $200$ GeV and data on $p^{\pm}n$ and $K^{\pm}n$
are available up to $\sqrt s \approx 25$ GeV.
Therefore, these reactions provide 
a good ground for investigating the effects of the
bounds obtained for the soft Pomeron intercept. In this
Section we present the results of global fits to the total cross
section data and discuss the implications in terms of factorization
and quark counting rules.

\subsection{Global fits to total cross sections}

Making use of
parametrization  (\ref{eq:total}) we perform  fits to $\sigma_{tot}$
data from  $p^{\pm} n$, $\pi^{\pm} p$, $K^{\pm} p$, $K^{\pm} n$, 
$\Sigma^{-} p$, $\gamma p$ and 
$\gamma \gamma$ scattering, exploring the effects of 
the extrema and constrained upper and lower bounds.

First, from Table 1, we fix the intercepts,
$\eta_{+}$ and $\eta_{-}$ at their central values and use
for $\epsilon$ the \textit{extrema bounds}, namely $\epsilon = 0.081$
and $\epsilon = 0.109$.
With this procedure we have 18 free parameters: the strengths
$X$, $Y_{+}$ and $Y_{-}$ for $p^{\pm} n$
$\pi^{\pm} p$, $K^{\pm} p$, $K^{\pm} n$
and  $X$, $Y_{+}$ for $\Sigma^{-} p$, $\gamma p$ and 
$\gamma \gamma$ scattering.
The total number of data points are 196 and therefore 
we have $F$ = 178.
The fit results for the extrema lower bound  were:

\begin{eqnarray}
& \sigma_{tot}^{p^{\pm}n} & = \ \ (21.02 \pm 0.19) \, s^{0.081} + (63.2 \pm 2.6) \, 
s^{-0.38}  \mp (15.19 \pm 0.82) \, s^{-0.42}, \nonumber \\ 
& \sigma_{tot}^{\pi^{\pm} p} & = \ \ (13.33 \pm 0.05) \, s^{0.081} + (25.18 \pm 0.60) \, 
s^{-0.38}  \mp (3.76 \pm 0.20) \, s^{-0.42}, \nonumber \\ 
& \sigma_{tot}^{K^{\pm} p} & = \ \ (11.92 \pm 0.06) \, s^{0.081} + (10.87 \pm 0.85) \, 
s^{-0.38}  \mp (6.85 \pm 0.23) \, s^{-0.42}, \nonumber \\
& \sigma_{tot}^{K^{\pm} n} & = \ \ (11.87 \pm 0.11) \, s^{0.081} + (10.0 \pm 1.5) \, 
s^{-0.38}  \mp (3.62 \pm 0.39) \, s^{-0.42}, \nonumber \\
& \sigma_{tot}^{\Sigma^{-} p} & = \ \ (20.13 \pm 0.53) \, s^{0.081} + 
(19.5 \pm 6.2) \, s^{-0.38}, \\
& \sigma_{tot}^{\gamma p} & = \ \ (0.067 \pm 0.001) \, s^{0.081} + (0.097 \pm 0.013) \, 
s^{-0.38}, \nonumber \\
& \sigma_{tot}^{\gamma \gamma} & = \ \ (0.00020 \pm 0.00001) \, s^{0.081} + 
(0.00016 \pm 0.00018) \, s^{-0.38}, \nonumber
\end{eqnarray}
with $\chi^2/F$ = 0.83. With the extrema upper bound we obtained

\begin{eqnarray}
 \sigma_{tot}^{p^{\pm} n} & = & \ \ (15.52 \pm 0.19) \, s^{0.109} + (55.7 \pm 1.7) \, 
s^{-0.28}  \mp (15.43 \pm 0.84) \, s^{-0.42}, \nonumber \\ 
\sigma_{tot}^{\pi^{\pm} p} & = & \ \ (10.13 \pm 0.05) \, s^{0.109} + (24.73 \pm 0.40) \, 
s^{-0.28}  \mp (3.82 \pm 0.21) \, s^{-0.42}, \nonumber \\
\sigma_{tot}^{K^{\pm} p} & = & \ \ (9.27 \pm 0.07) \, s^{0.109} + (13.85 \pm 0.57) \,
s^{-0.28}  \mp (6.96 \pm 0.23) \, s^{-0.42}, \nonumber \\
\sigma_{tot}^{K^{\pm} n} & = & \ \ (9.24 \pm 0.11) \, s^{0.109} + (13.25 \pm 0.99) \,
s^{-0.28}  \mp (3.67 \pm 0.39) \, s^{-0.42}, \nonumber \\
\sigma_{tot}^{\Sigma^{-} p} & = & \ \ (15.51 \pm 0.52) \, s^{0.109} + 
(25.1 \pm 4.0) \, s^{-0.28}, \\
\sigma_{tot}^{\gamma p} & = & \ \ (0.051 \pm 0.001) \, s^{0.109} + 
(0.105 \pm 0.009) \, s^{-0.28},  \nonumber \\
\sigma_{tot}^{\gamma \gamma} & = & \ \ (0.00015 \pm 0.00001) \, 
s^{0.109} + (0.00023 \pm 0.00011) \, s^{-0.28}, \nonumber 
\end{eqnarray}
with $\chi^2/F$ = 0.77.

These parametrizations for the $p^{\pm} n$, $\pi^{\pm} p$, $K^{\pm} p$, 
$K^{\pm} n$, $\Sigma^{-} p$, 
$\gamma p$ and $\gamma \gamma$ are displayed in Fig. 5 together with the 
experimental data.
We see that, with the exception
of the highest energy data points in $\pi^{-} p$,
$\gamma p$ and $\gamma \gamma$ scattering, the experimental data
are well described in both cases (upper and lower bounds). We note
that the highest points have the largest errors bars and, therefore,
little influence in global fits.
We shall return to this point in Sec. \ref{Sect:5}.

\begin{figure}[ht]
\begin{center}
\includegraphics[width=14.0cm,height=16cm]{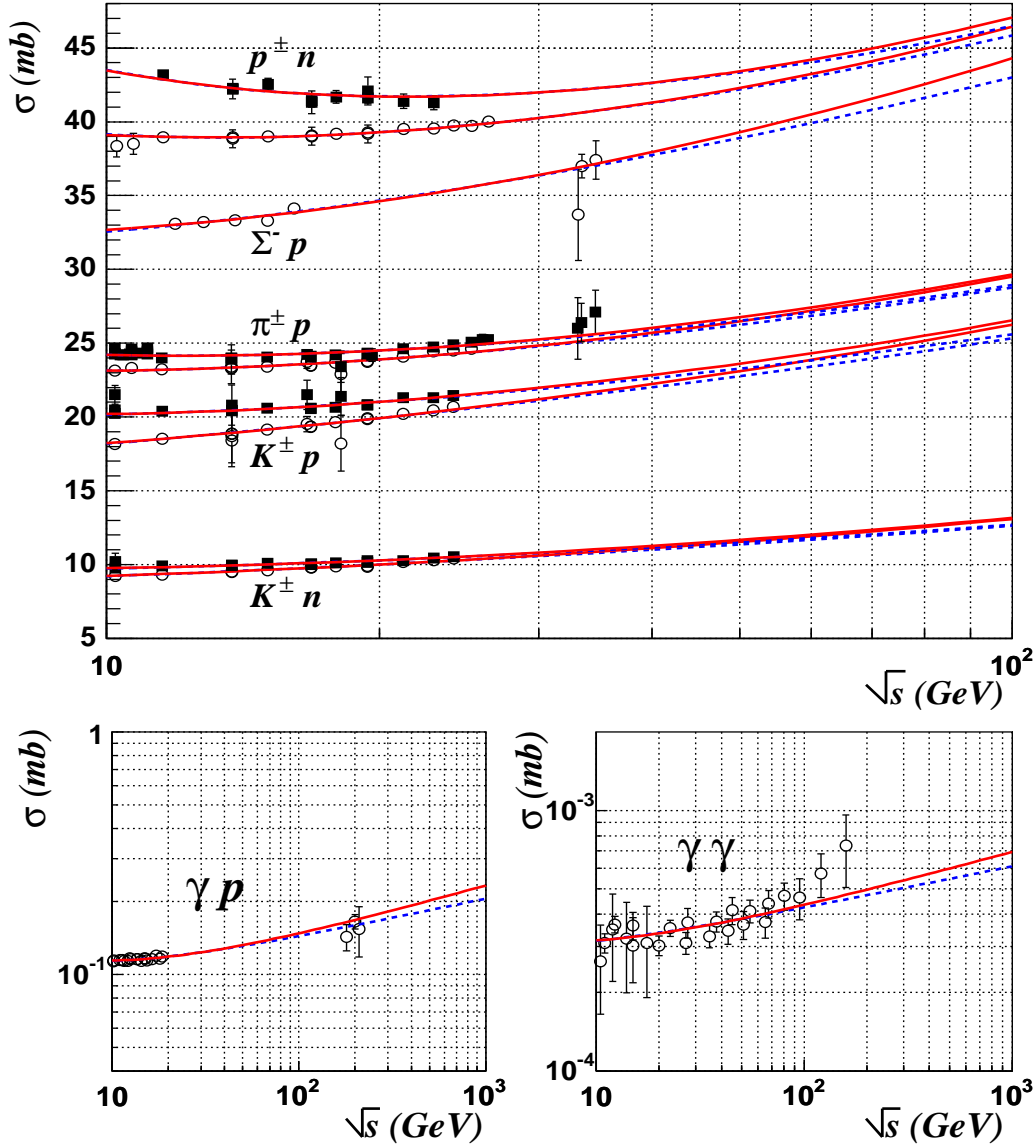}
\end{center}
\caption{Results of the global fits to proton-neutron, sigma-proton, pion-proton, 
kaon-proton, kaon-neutron,
gamma-proton and gamma-gamma total cross section data, with
fixed \textit{extrema bounds}: lower (dashed) and upper (solid), Eqs.
(13) and (14), respectively. For visualization, the $K^{\pm} n$ total cross
sections have been multiplied by a factor 0.5.} 
\label{fig5}
\end{figure}

Second, in order to check the effect of fixing the $C=+1$ and $C=-1$ 
trajectories to the spectroscopy data,
new fits to the $p^{\pm} n$, $\pi^{\pm} p$, $K^{\pm} p$, 
$K^{\pm} n$, $\Sigma^{-} p$, 
$\gamma p$ and $\gamma \gamma$ were performed using the values 
from Table \ref{tab:2}. For this
case, we also fixed the intercepts  $\eta_{+}$ and $\eta_{-}$ at their
central values and used for $\epsilon$ the 
\textit{constrained bounds}.

The fit results with the lower constrained bound  were:
\begin{eqnarray}
\nonumber
& \sigma_{tot}^{p^{\pm} n} & = \ \ (21.64 \pm 0.16) \, s^{0.080} + 
(82.4 \pm 3.3) \, s^{-0.452}  \mp (32.2 \pm 1.7) \, s^{-0.558}, \nonumber \\
& \sigma_{tot}^{\pi^{\pm} p} & = \ \ (13.71 \pm 0.04) \, s^{0.080} + 
(30.83 \pm 0.77) \, s^{-0.452}  \mp (7.72 \pm 0.42) \, s^{-0.558}, \nonumber \\
& \sigma_{tot}^{K^{\pm} p} & = \ \ (12.13 \pm 0.06) \, s^{0.080} + (13.2 
\pm 1.1) \, s^{-0.452}  \mp (14.40 \pm 0.48) \, s^{-0.558}, \nonumber \\
& \sigma_{tot}^{K^{\pm} n} & = \ \ (12.06 \pm 0.10) \, s^{0.080} + (12.12 
\pm 1.9) \, s^{-0.452}  \mp (7.73 \pm 0.82) \, s^{-0.558}, \nonumber \\
& \sigma_{tot}^{\Sigma^{-} p} & = \ \ (20.45 \pm 0.48) \, s^{0.080} + 
(24.4 \pm 8.0) \, s^{-0.452}, \\
& \sigma_{tot}^{\gamma p} & = \ \ (0.069 \pm 0.001) \, s^{0.080} + (0.119 
\pm 0.016) \, s^{-0.452}, \nonumber \\
& \sigma_{tot}^{\gamma \gamma} & = \ \ (0.00020 \pm 0.00001) \, s^{0.080} 
+ (0.00021 \pm 0.00024) \, s^{-0.452}, \nonumber
\end{eqnarray}
with $\chi^2/F$ = 0.70. With the upper constrained bound we obtained

\begin{eqnarray}
\nonumber
& \sigma_{tot}^{p^{\pm} n} & = \ \ (20.21 \pm 0.15) \, s^{0.089} + 
(89.6 \pm 3.2) \, s^{-0.452}  \mp (32.2 \pm 1.7) \, s^{-0.558}, \nonumber \\
& \sigma_{tot}^{\pi^{\pm} p} & = \ \ (12.83 \pm 0.04) \, s^{0.089} + 
(34.98 \pm 0.75) \, s^{-0.452}  \mp (7.63 \pm 0.42) \, s^{-0.558}, \nonumber \\
& \sigma_{tot}^{K^{\pm} p} & = \ \ (11.35 \pm 0.05) \, s^{0.089} + (16.9 
\pm 1.1) \, s^{-0.452}  \mp (14.39 \pm 0.48) \, s^{-0.558}, \nonumber \\
& \sigma_{tot}^{K^{\pm} n} & = \ \ (11.28 \pm 0.09) \, s^{0.089} + (15.9 
\pm 1.9) \, s^{-0.452}  \mp (7.71 \pm 0.82) \, s^{-0.558}, \nonumber \\
& \sigma_{tot}^{\Sigma^{-} p} & = \ \ (19.07 \pm 0.45) \, s^{0.089} + 
(31.7 \pm 7.8) \, s^{-0.452}, \\
& \sigma_{tot}^{\gamma p} & = \ \ (0.064 \pm 0.001) \, s^{0.089} + 
(0.139 \pm 0.016) \, s^{-0.452},  \nonumber \\
& \sigma_{tot}^{\gamma \gamma} & = \ \  (0.00019 \pm 0.00001) \, s^{0.089} 
+ (0.00031 \pm 0.00023) \, s^{-0.452}, \nonumber
\end{eqnarray}
with $\chi^2/F$ = 0.64. The corresponding curves are similar to those
in Fig. 5 with narrower limit regions and will not be displayed here.

It is worth to note that, differently from the $pp$ and $\bar{p}p$ case, 
fixing the
secondary Reggeon intercepts and reducing the Pomeron intercept produced a 
decrease on the $\chi^{2}/F$. Also, differently from the $pp$ and $\bar{p}p$
 case,
in all scenarios for the hadrons the $\chi^{2}/F$ remains lower than one, 
which seems to
indicate that these data, at least for their current energy range, are not
 quite sensitive 
to the behavior of the Pomeron and Reggeon intercepts.

\subsection{Factorization and quark counting}

The Pomeron contribution to the total cross section is characterized
by two parameters, the intercept $1 + \epsilon$ and the coefficient
$X$ (also referred to as strenght or coupling). In the previous sections we have
focused the discussion only on the investigation of bounds for the
intercepts (extrema and constrained). The analysis was based on the
selection of the fastest and slowest increase scenarios for the
total cross sections, allowed by the experimental data presently
available, and on fit procedures. Since in the fitting process both parameters,
$\epsilon$ and $X$, are statistically correlated (and also correlated with the
secondary Reggeon parameters), we expect that different bounds on
$\epsilon$ may, in some way, imply in different limit values for the
coefficients (see, for example, Tables 1 and 2).

In order to investigate the possible consequences and range of this
effect we consider here tests on two important properties related to
the strenght of the soft Pomeron: factorization and the additive-quark
rule. To that end, we first quickly review the conditions under which the
above properties are expected to hold \cite{ddln,bp} and then present
the numerical tests followed by a discussion on the obtained results.

According to the quark-counting rule the strenght of the soft Pomeron
to hadrons $A$ and $B$ is proportional to the number of valence
quarks inside each hadron, $N_A$, $N_B$: $X \propto N_A N_B$.
Therefore, from Eq. (4), if only a single Pomeron exchange dominates,
we expect that

\begin{equation}
{\sigma_{\mathrm{baryon}-\mathrm{baryon}} 
\over 
\sigma_{\mathrm{baryon}-\mathrm{baryon}}} \approx 1,
\qquad
{\sigma_{\mathrm{meson}-\mathrm{baryon}} 
\over 
\sigma_{\mathrm{baryon}-\mathrm{baryon}}}
\approx {2 \over 3},
\label{gc}
\end{equation}
meaning that the soft Pomeron couples to single quarks in a hadron, instead
of to the whole hadron.

In an elastic process mediated by the exchange of a single Pomeron,
the factorization of the coupling, as the product of the couplings at each vertex,
corresponds to the factorization of the associated residues functions,
$\gamma_{k}(t)$ in Eq. (2). This means that if we consider only the Pomeron
trajectory (or one Reggeon trajectory) the total cross sections for
different processes can be correlated since the same Pomeron coupling can
occur in different reactions. From Eqs. (1-3) and once one exchanged
trajectory is assumed, the total cross sections for elastic
processes like $p + p \rightarrow p + p$, $\gamma + \gamma \rightarrow
\gamma + \gamma$, and $\gamma + p \rightarrow \gamma + p$ should be related
by

\begin{equation}
\sigma_{p p} \, \sigma_{\gamma \gamma} \approx [\sigma_{\gamma p}]^2.
\label{f}
\end{equation}

Based on the above considerations, in order to test quark counting and
factorization we must consider only one contribution 
and therefore the leading one (Pomeron). Physically that means to investigate the region
of large enough energy where the contributions from the secondary reggeons can be 
neglected as compared with the Pomeron contribution. In other words,
we must assume that our parametrizations may be valid even beyond the
energy region with available data, so that the total cross section
in Eq. (4) can be expressed by

\begin{eqnarray}
\sigma_{tot}(s) \approx X s^{\epsilon}. \nonumber
\end{eqnarray}

With this formula and the values of the Pomeron coefficients from Tables 1 and 2
and Eqs. (15-18), we have calculated several ratios involving total cross sections
with the corresponding propagated errors. The results are displayed in Table 3
in a form so that both factorization and quark counting are fully verified for the
ratios equal to 1. We have the following comments on these results.

Roughly, the best agreements with (19) and (20)
have been obtained with the extrema bounds. As expected, the corresponding results with
the constrained bounds lie inside the error bars with the extrema bounds.

Concerning the quark counting rule (19), for the meson-baryon cross 
section we have that the ratio of the pion-proton to proton-proton 
cross section, obtained from the extrema bounds, agrees better with 
the expected value of $2/3$ for the lower $\epsilon$ value, although 
both results matched the expected ratio when the errors from the fit 
parameters are taken into account. For the kaon-proton reaction, the 
ratio to the proton-proton cross section does not agree with the $2/3$ 
value for both the extrema and constrained cases, falling some 15\% 
lower than that, on average. 
The same happens with the kaon-neutron to 
neutron-proton ratio, which is about 14\% lower, on average. However, 
it is advocated that the Pomeron coupling with strange quarks have 
a weaker strength as to light quarks \cite{ddln}, so that such deviations 
should be expected. 
We notice that for all the bounds the central 
values of the ratio $X_{K^{\pm} p} / X_{\pi^{\pm} p}$ are equal or 
greater than 0.88, which is a bit above the value 0.87 obtained 
by Donnachie and landshoff with degenerate Reggeon trajectories \cite{dl92} .
We also notice that the kaon-neutron to kaon-proton ratio show a very 
stable pattern, been equal to unity, or very close to that, for all 
bounds considered. As for the baryon-baryon cross section, we have that 
the agreement between the ratio of neutron-proton to proton-proton and 
the expected unit value from the quark counting rule is quite noticeable, 
for all $\epsilon$. The other baryon-baryon reaction available,
sigma-proton, provides a ratio to proton-proton that is systematically 
lower than one, in agreement with the hypothesis of a weaker coupling 
of the Pomeron with strange quarks mentioned above.

Concerning the factorization rule (20), both constrained and extrema upper
bounds provided ratios closer to 1 than the corresponding lower bounds.
Despite the fact that, taking into account the error bars, the only result
consistent with the value 1 had been obtained with the extreme upper
bound, roughly, we can say that the factorization rule is barely verified in all
the  cases.

We conclude that, in the context of the procedure used, the effects of the
extrema bounds are not in disagreement with what is generally expected from
both quark counting and factorization rules.

\begin{table}[h]
\begin{center}
\caption{Results for the quark counting rule and factorization
using the Pomeron strengths obtained with both
extrema and constrained upper and lower bounds.}
\begin{tabular}{ccccc} 
\hline
Bounds: & \multicolumn{2}{c}{extrema} & \multicolumn{2}{c}{constrained} \\
 & lower & upper & lower & upper \\
 $\alpha_{\tt I\!P}(0)$: & 1.081 & 1.109 & 1.080 & 1.089   \\ 
\hline
$ \sigma_{n p} \over  \sigma_{pp}$ & 1.03$\pm$0.04 & 0.94$\pm$0.07 
& 1.02$\pm$0.02 & 0.99$\pm$0.02  \\

$ \sigma_{\Sigma^{-} p} \over  \sigma_{pp}$ & 0.98$\pm$0.05 & 
0.94$\pm$0.08 & 0.96$\pm$0.03 & 0.93$\pm$0.03  \\

$ {3 \over 2} {\sigma_{\pi^{\pm} p} \over \sigma_{pp}}$ & 0.98$\pm$0.04
 & 0.93$\pm$0.07 & 0.96$\pm$0.01 & 0.94$\pm$0.01 \\

$ {3 \over 2} {\sigma_{K^{\pm} p} \over \sigma_{pp}}$ & 0.87$\pm$0.04 &
 0.85$\pm$0.06 & 0.85$\pm$0.01 & 0.83$\pm$0.01   \\

$ {3 \over 2} {\sigma_{K^{\pm} n} \over \sigma_{np}}$ & 0.85$\pm$0.04 &
 0.90$\pm$0.07 & 0.84$\pm$0.01 & 0.84$\pm$0.01   \\

$ {\sigma_{K^{\pm} n} \over \sigma_{K^{\pm} p}}$ & 1.00$\pm$0.04 & 1.00$\pm$0.08
 & 0.99$\pm$0.02 & 0.99$\pm$0.02   \\

$ {\sigma_{K^{\pm} p} \over \sigma_{\pi^{\pm} p}}$ & 0.89$\pm$0.006 & 0.92$\pm$0.008 
& 0.88$\pm$0.005 & 0.88$\pm$0.005   \\

$ \sigma_{p p} \, \sigma_{\gamma \gamma} \over (\sigma_{\gamma p})^{2}$
 & 0.91$\pm$0.06 & 0.96$\pm$0.09 & 0.91$\pm$0.04 & 
0.93$\pm$0.04  \\ 
\hline
\label{tab:3}
\end{tabular}
\end{center}
\end{table}

\section{\label{Sect:5}Conclusions and final remarks}

Disagreements between different experimental estimations of the total
cross sections at the highest energies ($pp$ and $\bar{p}p$ scattering),
lead to uncertainties in the determination of the soft Pomeron
intercept. By means of an extended Regge parametrization
(non-degenerated meson trajectories) the effect of these discrepancies
has been translated into the estimation of extrema bounds for the
intercept \cite{lm03}. In this work, we have investigated the consequences
of these bounds in the study of the total cross sections from 
$p^{\pm}n$, $\pi^{\pm}p$, $K^{\pm}p$, $K^{\pm}n$,
$\Sigma^{-}p$, $\gamma p$ and $\gamma \gamma$ scattering and have presented
tests on factorization and quark counting rules. The effects of secondary
Reggeon constraints from Chew-Frautschi plots (spectroscopy data) have also 
been treated.

We have shown that both extrema bounds,
1.109 and 1.081 lead to good
descriptions of the bulk of the experimental data presently available
on total cross sections for all the reactions investigated. 
Tests on factorization and quark counting indicate that the results
with all the bounds are not in disagreement with what is 
generally expected
from standard fits to scattering data.
This means that, if we consider the average of the extrema bounds, 
$\alpha_{\tt I\!P}(0) = 1.095 \pm 0.020$,
the experimental data investigated are not sensitive to typical 
uncertainties of 2\%
in the value of the  intercept. 
We understand that this conclusion 
brings novel 
information on the numerical interval that could be associated with
a truthful value for the soft Pomeron intercept.
However, extrapolation beyond
the regions with available data indicate different scenarios for the
rise of the total cross sections. Certainly, future data might select the 
best bound.

The constrained bounds imposed by the fitted secondary trajectories 
(Chew-Frautschi plots)
reduce the previous intercept bounds, allowing the estimation
$\alpha_{\tt I\!P}(0) = 1.085 \pm 0.006$ (average) .
From Fig. 4 the constrained bounds 
show qualitative agreement with a slower increase scenario for the total cross
section, as represented by Ensemble II + BHS (the E710 and E811 results for
$\bar{p}p$ and the results by Block, Halzen and Stanev for $pp$ at
cosmic-ray energies). However, it should be noted that the best statistical
results were obtained in the case of the extrema bounds, as indicated 
in Tables 1 and 2.

We note that, as in other approaches, the highest data points from $\pi^{-} p$,
and $\gamma \gamma$ scattering are not described (Fig. 5). 
Although the extrapolations
predict different behaviors for $\sigma_{tot}(s)$, neither bound are
able to describe the above points in a reasonable way, specially in the
$\gamma \gamma$ scattering. This last case may suggest the necessity of an
additional component, as the hard Pomeron \cite{land01}.

In this paper we have considered a specific extended parametrization
characterized by degenerate secondary trajectories $a_2/f_2$ and 
$\rho / \omega$, Eq. (4). Another aspects that may affect the value of the bounds 
and even the minimal energy for stable fits are the possibility of
nondegenerate $\rho$ and $\omega$ trajectories and also a non-linear $f_2$ 
trajectory. We are presently investigating these aspects with both
scattering and spectroscopy data. 

\begin{ack}
We are thankful to FAPESP for financial
support (Contract N. 00/00991-7 and 00/04422-7).
\end{ack}

\end{document}